\documentclass[letters,usenatbib]{mnras}
\usepackage{newtxtext,newtxmath}
\usepackage[T1]{fontenc}

\usepackage{fix-cm}
\DeclareRobustCommand{\VAN}[3]{#2}
\let\VANthebibliography\thebibliography
\def\thebibliography{\DeclareRobustCommand{\VAN}[3]{##3}\VANthebibliography}

\usepackage{graphicx}	
\usepackage{amsmath}	

\defcitealias{Gurkan2018}{G18}	
\defcitealias{2023MartinH_optcounterparts}{H23}	
\defcitealias{S21}{S21}
\defcitealias{sdas2024}{D24}	
\defcitealias{2026Shenoy}{S26}
\title[Every galaxy is a radio source]{Every galaxy is a radio source: a stellar-mass dependent minimum 150\,MHz luminosity identified in the LoTSS Deep Fields}
\author[S.\,Shenoy et al.]{Shravya\,Shenoy,$^{1}$\thanks{E-mail: s.shenoy@herts.ac.uk}
Daniel\,J.\,B.\,Smith$^{1}$, Martin\,J.\,Hardcastle$^{1}$,
Soumyadeep Das$^{1}$, Marina I. Arnaudova$^{2,4}$, \newauthor 
Kenneth\,J.\,Duncan$^{2}$,  Luke R. Holden$^{1}$,
Huub J. A. R\"ottgering$^{3}$
\\
$^{1}$Centre for Astrophysics Research, University of Hertfordshire, Hatfield, AL10 9AB, UK\\
$^{2}$Institute for Astronomy, University of Edinburgh, Royal Observatory, Blackford Hill, Edinburgh, EH9 3HJ, UK.\\
$^{3}$Leiden Observatory, Leiden University, PO Box 9513, 2300 RA Leiden, The Netherlands\\
$^{4}$U.S. National Science Foundation National Optical-Infrared Astronomy Research Laboratory, 950 North Cherry Avenue, Tucson, AZ 85719, USA
}

\date{Accepted XXX. Received YYY; in original form ZZZ}

\pubyear{\the\year{}}

\begin{document}
\label{firstpage}
\pagerange{\pageref{firstpage}--\pageref{lastpage}}
\maketitle

\begin{abstract}
The second data release of the LOFAR Two-metre Sky Survey (LoTSS) Deep Fields includes the deepest wide-field 150\,MHz mosaics in existence, covering the ELAIS-N1 field. Combining these radio observations with star-formation rate (SFR) and stellar mass estimates for 113,228 near-infrared selected galaxies derived using \texttt{Prospector}, we re-examine the relationship between galaxies' 150\,MHz luminosities ($L_{150\rm MHz}$), SFRs and stellar masses. We not only recover the established radio-luminosity–SFR relation and its dependence on stellar mass, but also discover a mass-dependent minimum radio luminosity that is independent of SFR. This behaviour is especially evident in galaxies with low specific SFR (sSFR $\equiv$ SFR/$M_*$), at $L_{150\rm MHz}\sim10^{20}-10^{22}$~W~Hz$^{-1}$, where star-forming galaxies are expected to dominate. The non-stellar emission is consistent with low-level accretion onto supermassive black holes (i.e. weak active galactic nuclei, AGN), spanning effective Eddington-scaled accretion rates $\lambda_{\rm{Edd}}\sim10^{-5}$–$10^{-4}$, suggesting that radio emission due to accretion is widespread and effectively continuous. A two-component model combining star formation- and accretion-driven emission reproduces the observed trends, demonstrating that AGN activity can bias radio-based SFR estimates. The low radio luminosities probed in this work provide a first glimpse of the populations that will be routinely accessible with future facilities such as LOFAR~2.0 and the Square Kilometre Array.
\end{abstract}

\section{Introduction}
In galaxies where stellar processes are the primary contributors to radio emission (e.g. \citealt{1954Baade, Condon1992}), the radio continuum arises mainly from two mechanisms: thermal (free–free) emission produced by ionised gas in H\,II regions around massive, young stars, and non-thermal (synchrotron) emission generated by relativistic electrons accelerated in (Type II) supernova shocks. Both mechanisms are directly associated with populations of short-lived stars, making radio emission a reliable, though indirect, tracer of recent star formation. This reliability is further supported by the well-established far-infrared–radio correlation (FIRC; e.g. \citealt{1971Vanderkruit, 1985DeJong, 2001Yun, Bell2003, 2014smithfirrc, Read18, 2019Wang_FIRC, Delvecchio21, 2024Cook}), in which far-infrared emission is assumed to predominantly result from dust heated by the same population of massive stars that produces the supernovae.\\
\\
A relationship between radio luminosity and star formation rate is well established for star-forming galaxies (SFGs) across a wide range of radio frequencies (e.g. \citealt{Bell2003, 2009Garn, 2015Boselli,Brown2017, 2017Rivera, Davies2017, Gurkan2018, 2020Leslie,2025Tabatabei, 2026Seymour}). There is also growing evidence for non-linearity at low frequencies in the $L_\mathrm{150MHz}$--SFR relation, including dependence on stellar mass (e.g. \citealt{Gurkan2018}, henceforth \citetalias{Gurkan2018}; \citealt{Read18}; \citealt{S21}, henceforth \citetalias{S21}; \citealt{2022McCheyne}; \citealt{2022Heesen}; \citealt{sdas2024}, henceforth \citetalias{sdas2024}) which may require additional physics (e.g. \citealt{2019Panessa}). Hypotheses for the origin of the mass dependence include calorimetric arguments \citep{NickBeck1997, 2010LackiThompson}, Type\,Ia supernovae (\citealt{1984Sramek, 2023Kool_type1a}), X-ray binaries, millisecond pulsars \citep{Sudoh21}, and galactic wind-ISM interaction (e.g. \citealt{1971Mathews_winds, 2024Todd_winds}).\\
\\
However, caution is required, since the likelihood of a galaxy hosting an active galactic nucleus (AGN), particularly a radio AGN, increases sharply with stellar mass (e.g. \citealt{2005Best, 2012Aird, 2019Sabater, 2025Kondapally}), raising concerns about contamination of radio emission in massive systems. Low-level AGN activity, even in otherwise quiescent galaxies, can result in excess radio luminosity that does not correlate with SFR, and may complicate the use of radio emission as a straightforward tracer of star formation in massive or evolved systems \citep[e.g.][]{2008Hodge}.\\
\\
\citetalias{S21} and \citetalias{sdas2024} used 150\,MHz observations from the first data release (DR1) of the LOFAR Two-metre Sky Survey Deep Fields (LoTSS; \citealt{2021Sabater,2021Duncan_photoz_and_mass,2021Kondapally,2021Tasse})  to confirm the existence of stellar-mass-dependent trends in the $L_{150\rm MHz}$--SFR relation \citep[which had previously been proposed over the \textit{Herschel} Astrophysical TeraHertz Large Area Survey North Galactic Pole field, the \textit{H}-ATLAS NGP, by][]{Gurkan2018}, although recent works by \citealt{2026Shenoy} (henceforth \citetalias{2026Shenoy}) and \citet{2026jackson} have suggested that the mass dependence may be weaker than previously identified. In this context, new ultra-deep 150\,MHz maps of the European Large Area ISO Survey (ELAIS)-N1 deep field from the second data release (DR2) of the LoTSS Deep Fields \citep{2025Shimwell_LoTSSDR2Deep} now provide the deepest 150\,MHz observations available, enabling the radio properties of galaxies to be studied at flux densities that will become routinely accessible in the era of LOFAR~2.0 \citep{lofar2.0} and SKA \citep{2009SKA}. \\
\\
These factors now motivate a closer examination of the role of stellar mass in the $L_{150\rm{MHz}}$--SFR relation.  Building on the framework developed by \citetalias{S21} and \citetalias{sdas2024}, in this letter, we use these new deeper radio data to revisit the $L_{150\rm{MHz}}$--SFR relation. The structure of this work is as follows: Section \ref{sec_data} details the datasets and sample selection criteria. Section \ref{results} presents an updated investigation of the relation between $L_{150\rm MHz}$ and SFR and its dependence on stellar mass. In Section \ref{discussion} we present a discussion of our key results and in Section \ref{conc} we present the conclusions. Throughout this work we adopt a flat $\Lambda$CDM cosmology with $H_0 = 70$~km~s$^{-1}$~Mpc$^{-1}$, $\Omega_m = 0.3$, and $\Omega_\Lambda = 0.7$ and a constant spectral index value of $\alpha = 0.7$, assuming the convention that $S_{\nu} \propto \nu^{-\alpha}$. 

\vspace{-0.5cm}
\section{Data}\label{sec_data}
In this work, we combined the deepest 150\,MHz observations from LoTSS Deep Fields DR2 \citep{2025Shimwell_LoTSSDR2Deep}, which cover the ELAIS-N1 deep field \citep{2000Oliver_ELAIS}, with star formation rates and stellar masses derived from spectral energy distribution (SED) fitting of aperture-matched multi-wavelength photometry (see \citealt{2021Kondapally} for details regarding the catalogue) using the \texttt{Prospector} code \citep{2021Johnsonprospector}, as described in \citetalias{sdas2024}.
\subsection{Radio observations - LoTSS Deep Fields DR2}\label{radio_observations}
We use the LOw Frequency ARray (LOFAR; \citealt{VanHaarlem13}) Two-metre Sky Survey Deep Fields Data Release 2 (LoTSS Deep DR2), centred at 150\,MHz. 24.53 square degrees of the field centered around  RA=16:11:00 and Dec=+55:00:00 (J2000) were observed with the LOFAR High Band Antenna (HBA) for a total of 505 hours over 7.5 years. The data reach a root mean square (RMS) noise level of 10.7~$\mu$Jy beam$^{-1}$ at the centre of the image at a spatial resolution of 6$''$ \citep{2025Shimwell_LoTSSDR2Deep}. The DR2 maps reach an average RMS $\sim$0.68 times that of the DR1 data \citep{2021Sabater}, enabling substantially improved sensitivity to faint radio emission.

\subsection{Sample selection}\label{sampleselection}

Following \citetalias{S21}, we begin with sources brighter than 10$~\mu$Jy in 3.6~$\mu$m data from the \textit{Spitzer} Wide-Area Infrared Extragalactic Survey \citep[SWIRE;][]{2003SWIRELonsdale}, corresponding to a signal-to-noise ratio (S/N) of $\geq 5$. This selection benefits from both the depth of these data and their insensitivity to dust obscuration. This sample contains 195,324 $z_\mathrm{best} < 1$ sources, where $z_\mathrm{best}$ comes from the LoTSS Deep fields DR1 value-added catalogue and corresponds to a reliable spectroscopic redshift if available, and a photometric redshift from \citet{2021Duncan_photoz_and_mass} otherwise. We include only those sources with the most reliable photometric redshifts\footnote{\cite{2021Duncan_photoz_and_mass} demonstrated that, for host-dominated galaxies, photometric redshift estimates have robust scatter with a normalised median absolute deviation $\sigma_{\rm{NMAD}} < 0.04 (1+z)$ in the EN1 field for $z < 1$; we select this sample by requiring \texttt{z1\_max $-$ z1\_min > 0.02}. }, and by considering galaxies regardless of their $S_{\rm{150\,MHz}}$ S/N we avoid a potential source of luminosity bias (see e.g. Appendix~B of \citetalias{S21}). 174,742 sources from the parent sample have stellar mass and SFR estimates obtained via SED fitting with \texttt{Prospector} (\citetalias{sdas2024}). From this subset, we further selected only those with coverage in both Pan-STARRS and UKIDSS, and excluded sources located near optically bright stars to obtain robust photometry (see \citealt{2021Kondapally} for details), to get 130,565 sources. Lastly, to mitigate the influence of extended radio emission larger than the LoTSS 6$''$ beam, we include only those sources with $z_\mathrm{best} > 0.05$\footnote{Our results are consistent if we choose a more restrictive $z_\mathrm{best} > 0.1$}, leaving 126,133 sources in our sample.\\
\\
We proceed with 126,127 galaxies where we believe that the \texttt{Prospector} SED fits are working well, requiring $7 \leq \log(M_*/M_{\odot}) \leq 12$ and $-3\,\leq \log(\psi/M_{\odot}\rm{yr}^{-1})\,\leq3$ where $\log(M_*/M_{\odot})$ and $\log(\psi/M_{\odot}\rm{yr}^{-1})$ represent the \texttt{Prospector} median-likelihood values of stellar mass and SFRs, respectively. We remove known AGN based on a number of criteria, using the flags from \cite{2021Duncan_photoz_and_mass} to remove galaxies that are classified as AGN-dominated, detected either using optical spectroscopy, mid-infrared colour cuts (based on \citealt{2012Donley_midIR_flag}), X-ray luminosity cuts based on data from the Second ROSAT All-Sky Survey \citep{2016Boller_xray_agn} and the XMM-Newton Slew Survey\footnote{\url{https://www.cosmos.esa.int/web/xmm-newton/xmmsl2-ug}}, or those included in the Million Quasar Catalogue \citep{2015Flesch_millquasar}, reducing the sample to 125,403\,galaxies. \\
\\
Approximately 15 per cent of these 125,403 galaxies have optical spectroscopy available from the first data release of the Dark Energy Spectroscopic Instrument (DESI DR1; \citealt{2024DESIEDR}). \cite{2025Arnaudova} performed probabilistic spectroscopic classifications of 4,471 of these LoTSS sources at $z < 0.98$, updating the method used for $z < 0.46$ galaxies from \cite{drake2024}. Adopting a confidence threshold of 90 per cent to remove the spectroscopically identified AGN, our sample reduces to 124,703 galaxies. We also exclude 321 sources\footnote{Of these, 240 sources are classified as radio excess based on equation 1.} with probability $\mathcal{P}$(RadioExcess)\,$>0.9$, i.e., corresponding to galaxies whose radio luminosity is significantly higher than expected from their Balmer-corrected H$\alpha$ luminosity. Such sources are therefore unlikely to be purely star-forming galaxies, leaving a sample of 124,382 sources. As these probabilistic classifications are only available for the LoTSS-detected sources, for the remainder of the spectroscopic sample, we identify AGN using the nebular emission-line fluxes from the \textsc{fastspec}\footnote{\url{https://fastspecfit.readthedocs.io/en/latest/}; Iron VAC} catalogue (\citealt{2023Moustakas}, J. Moustakas \textit{in prep.}) via the BPT \citep[e.g.][]{BPT1981, 1987Veilleux, 2003Kauffmannbpt, 2006Kewley} and Mass-Excitation \citep[MEx;][]{2011Juneau} diagrams. We also use the relation between $L_{\mathrm{H}\alpha}$ and $L_{150\,\mathrm{MHz}}$ from \cite{2025Arnaudova} (equation 1) to remove radio-excess galaxies. Together, these cuts reduce the sample to 118,823 galaxies. This removes any detectable radiatively efficient AGN, although weak AGN may remain.\\
\\
For the remaining 85 per cent of the sample, in the absence of H$\alpha$ fluxes, we rely on the assumption that the correlation between SFR and $L_{150\rm{MHz}}$ provides a reliable prediction of the radio emission expected purely from stellar processes and that sources with radio luminosities exceeding this relation by more than 3$\sigma$ can be flagged as having significant AGN contribution (\citetalias{S21}; \citealt{2023Best}; \citetalias{sdas2024}). To identify and remove these radio-excess sources from our sample, we use the mass-independent form of the $L_{150\rm MHz}$--SFR relation from \citetalias{sdas2024}, given by:
\begin{equation}
    \log_{10}\,(L_{\rm{150~MHz}}/\rm{W~Hz}^{-1}) = 22.024 + 1.019 \log_{10}\,(\psi/M_{\odot}yr^{-1})
\end{equation}
where $\psi$ represents the SFR of the galaxy. We remove galaxies that lie more than 1 dex ($\approx 3\sigma$) above the relation, resulting in 116,176 galaxies.\footnote{We note here that using the stellar-mass-dependent relation from \citetalias{sdas2024} to remove radio excess sources results in the removal of 2,185 sources instead of 2,647 sources and that this does not affect our key results in Section \ref{results}.}
We also remove 2,948 galaxies identified as hosting radiative AGN in the \texttt{Prospector} SED fits.\footnote{Specifically, those with $f_{\rm AGN,16} > 10^{-3}$; see \citetalias{sdas2024} for details.} This gives us a sample of 113,228 galaxies.\\
\\
To remove compact radio AGN from our sample, we rely on a selection method based on the principle that radio surface brightness and brightness temperature ($T_b$; \citealt{Condon1991}) above a threshold cannot be explained by stellar processes alone. For sources in the ELAIS-N1 field with sub-arcsec imaging available from \cite{2024JurjendeJongILTElaisN1}, \cite{2025Morabito} have combined arcsec- and sub-arcsec-resolution imaging to identify such a subsample. Using the threshold recommended in their work, we identify 57 galaxies ($\sim 0.05$ per cent of our sample of 113,228 galaxies) which have a significant AGN contribution to the 150\,MHz flux densities. We verify that excluding these galaxies does not change our results, but note that this could change in the future with the identification of more such sources using intermediate-resolution radio data and deeper radio data from the LOFAR Ultra-Deep Observations (LUDO) survey (Best et al. \textit{in prep}.).\\
\\
For the 11 per cent of these 3.6~$\mu$m sources that are identified as the hosts of 150\,MHz sources in the LoTSS catalogue \citep{2025Shimwell_LoTSSDR2Deep}, we use the radio flux densities measured using \texttt{PyBDSF} \citep{2015pybdsf}. For the rest, following \citetalias{sdas2024} and \citetalias{S21}, we use the peak pixel flux value (in Jy beam$^{-1}$) from the primary beam-corrected LoTSS Deep DR2 mosaic, which represents the maximum likelihood flux estimate for a source, assuming that it is unresolved on the scale of the LoTSS beam (e.g. \citetalias{S21}). Similarly, the radio flux density uncertainties are either adopted from the catalogue, or measured from the corresponding pixels in the LoTSS \texttt{PyBDSF} RMS mosaics otherwise\footnote{\url{https://lofar-surveys.org/deepfields_dr2.html}}. Compared to the samples of \citetalias{S21} and \citetalias{sdas2024}, our sample contains 51.8 per cent more sources with radio flux density S/N $> 3$ (36,536 versus 24,062), owing to the deeper radio observations. Overall, 32.3 per cent of the sources in our sample have radio flux densities measured at S/N $>3$ and 15.1 per cent have S/N $> 5$. To assess the reliability of our pixel flux and uncertainty estimates, we repeated the pixel photometry procedure at 5,000 randomly selected sky positions, where no real sources are expected. The resulting distribution of signal-to-noise ratios closely follows a standard normal (Gaussian) distribution, indicating that the noise properties are well-behaved and that our measurement uncertainties are well-characterised.

\section{Results}\label{results}
\begin{figure*}
    \centering
    \includegraphics[width=\linewidth, trim=0cm 0.4cm 0cm 0cm]{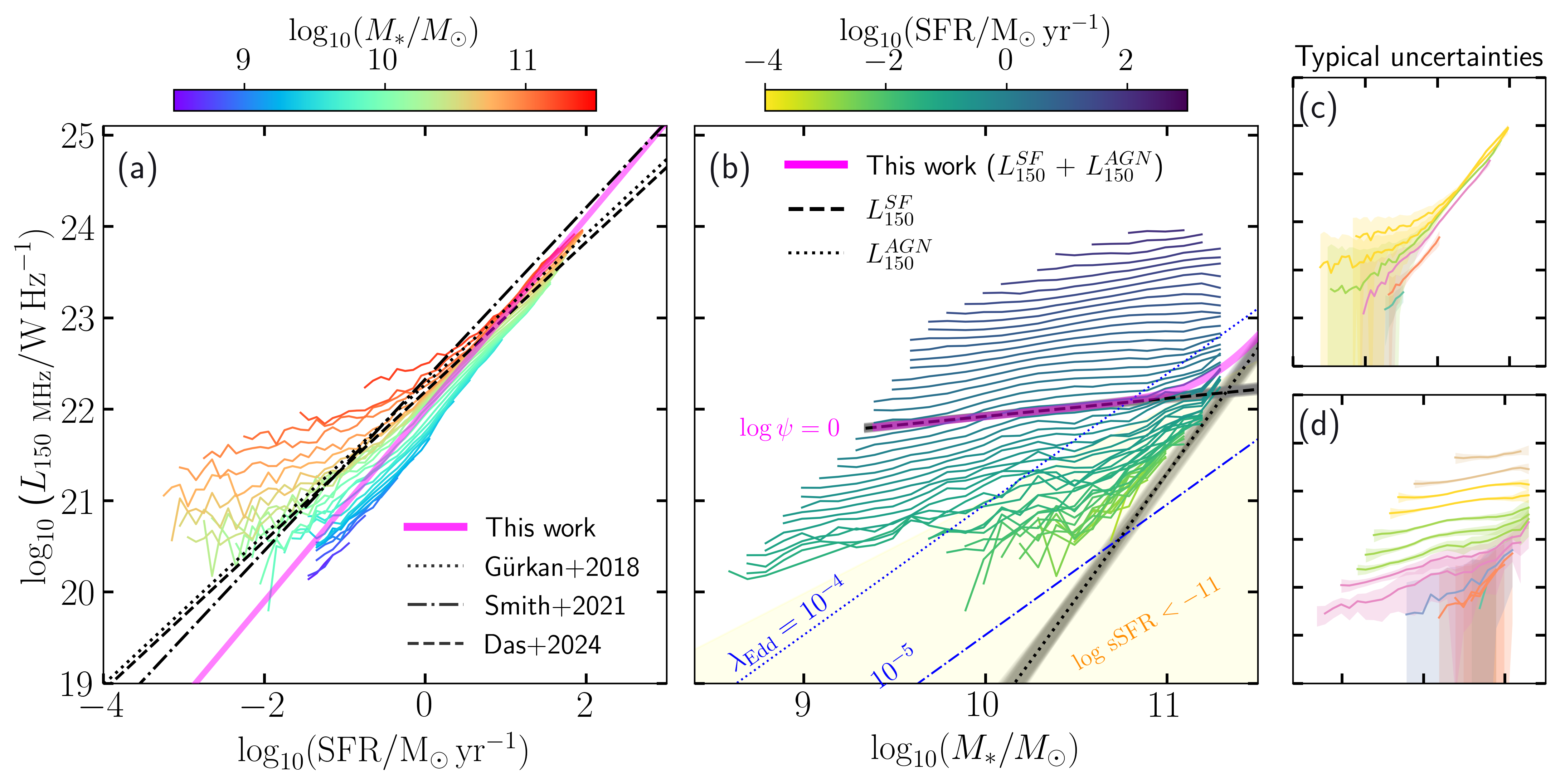}
    \caption{\textit{Panel (a):} Median radio luminosity, $\log(L_{\rm 150\,MHz}/\rm W\,Hz^{-1})$, as a function of SFR, with each line corresponding to a stellar mass bin, as indicated by the colour bar on the top. Overplotted are the best-fitting relations from \citetalias{Gurkan2018}, \citetalias{S21}, and \citetalias{sdas2024}, as well as the bias-corrected one from this work (in pink). \textit{Panel (b):} Median radio luminosity as a function of stellar mass, with line colour indicating SFR, as shown by the colour bar. The yellow shaded region denotes $\log \mathrm{sSFR / yr^{-1}} \leq -11$. The blue dotted and dash-dotted lines indicate the effective Eddington-scaled accretion rates of SMBHs contributing to the radio luminosity, calculated using the relation reported by \citet{2010Cavagnolo}, as described in Section \ref{discussion}. Indicative radio luminosities calculated using equations \ref{SFmodel} ($L_{150}^{\rm{SF}}$) and \ref{agneq} ($L_{150}^{\rm{AGN}}$) are shown by the black dashed and dotted lines respectively and the pink line corresponds to the sum of $L_{150}^{\rm{SF}}$ at an indicative $\log_{10} \psi = 0$ and the mass dependence of $L_{150}^{\rm{AGN}}$, as given by equation \ref{maineq}. In panels (c) and (d), we show the uncertainties on the median $L_{150}$ for a representative subset of the stellar mass and SFR bins respectively, as shaded regions using a different colour palette to aid visibility. The axes and axis ticks in panels (c) and (d) match those of panels (a) and (b) respectively. }
    \label{fig:mass_dep}
\end{figure*}

Motivated by the finding of \citetalias{2026Shenoy} that SFR and stellar mass are the primary predictors of the radio luminosity of star-forming galaxies, we first visualise this dependence by computing the median radio luminosity in bins of SFR and stellar mass. We do this by taking into account the asymmetric uncertainties on values of SFR and stellar mass, following the method outlined in \citetalias{S21}. In panel (a) of Fig.~\ref{fig:mass_dep}, we plot the median likelihood values of radio luminosity as a function of SFR at fixed stellar mass, with stellar mass indicated by the colour bar located above. Similarly, in panel (b), we plot the median likelihood values of radio luminosity as a function of stellar mass at fixed SFR, indicated by the colour bar. Typical uncertainties associated with the median $L_{150}$,  are shown for a representative subset of lines in panel (c) and (d) of the same figure, with axes identical to panels (a) and (b), respectively.
The median bin occupancy is 61 galaxies (minimum 15), with a typical $L_{150}$ S/N of $\sim$6.6$\sigma$, though as panels (c) and (d) show, this varies significantly. 
The $L_{150\,\rm MHz}$--SFR relation exhibits a dependence on stellar mass similar to what has been reported in previous studies, as shown in the panel (b) of Fig.~\ref{fig:mass_dep}. However, we also see that at $\log_{10}(M_*/M_{\odot})\,\geq\,10.5$ and $\log_{10}(L_{150\rm MHz}/\rm~W~Hz^{-1}) < 22$, the radio luminosities of galaxies no longer scale with SFR, but instead converge to a value which depends only on the stellar mass; it is important to note that this is in addition to the mass dependence trends found in SF-related radio emission  by \citetalias{Gurkan2018}, \citetalias{S21}, and \citetalias{sdas2024}, and consistent with mass-dependence found in the radio-loud AGN population (e.g. \citealt{2019Sabater}). The sensitivity and depth of the LoTSS Deep Fields DR2 enable us to disentangle these components, demonstrating that the AGN-related mass-dependent contribution is detectable even at radio luminosities where SFGs dominate. Given that this behaviour occurs for the most quiescent galaxies\footnote{A $\log$ (sSFR / yr$^{-1}$) $< -11$ threshold has been widely used in the literature to identify quiescent galaxies \citep[e.g.][]{2013Ilbert, 2023Houston}.}, i.e. those with low sSFR, the radio emission likely reflects residual low-level AGN activity, which becomes increasingly prevalent at higher stellar mass \citep[e.g.][]{2019Sabater,2025Kondapally}. To characterise the observed behaviour, we model the median $L_{150\, \mathrm{MHz}}$ as the sum of a star-formation (SF) component and a mass-dependent term, which may arise from low-level AGN emission:
\begin{equation}\label{maineq}
L_{150\, \mathrm{MHz}} = L_{150}^{\rm SF} + L_{150}^{\rm AGN}.
\end{equation}
For the star-formation component, we adopt the mass-dependent form used in previous works:
\begin{equation}\label{SFmodel}
L_{\textrm{150\,MHz}}^{\rm SF} = 10^{L_{\rm C1}} \left(\frac{\psi}{{M}_{\odot} \rm{yr}^{-1}}\right)^{\beta} \left( \frac{M_*}{10^{10} {M}_{\odot}} \right)^{\gamma};
\end{equation}
while the AGN component is modelled as an additional mass-dependent term for which we adopt a similar expression:
\begin{equation}\label{agneq}
L_{\textrm{150\,MHz}}^{\rm AGN} = 10^{L_{\rm C2}} \left( \frac{M_*}{10^{11} {M}_{\odot}} \right)^{\delta}.
\end{equation}
We normalise the stellar mass in the AGN component to $10^{11}\,M_{\odot}$, in contrast to the $10^{10}\,M_{\odot}$ normalisation adopted for the SF component, as the former corresponds to the characteristic mass scale where low-level AGN activity is observed to become significant in galaxies \citep{2019Sabater}, consistent with the trend seen in our data (panel b in Fig.~\ref{fig:mass_dep}).\\
\\
The free parameters of the model are $\{L_{\rm C1}, \beta, \gamma, L_{\rm C2}, \delta\}$. We fit these following the method from \citetalias{2026Shenoy}. To summarise briefly, best-fitting values are obtained through an initial optimisation, followed by Markov Chain Monte Carlo (MCMC) sampling with the \texttt{emcee} ensemble sampler. We use 32 walkers and run each chain for 5,000 steps, discarding an initial burn-in. The priors on $L_{\rm{C1}}, \beta$ and $\gamma$ are the same uniform priors as those used for equation \ref{SFmodel} in previous works \citepalias{S21, sdas2024}. For the AGN component, we impose uniform priors between $18 < L_{\rm C2} < 25$ and $0 < \delta < 4$, allowing for a wide range of normalisations and stellar-mass dependencies. Parameter combinations falling outside these bounds are assigned zero prior probability.  We also use bespoke mock simulations similar to those in \citetalias{2026Shenoy} to perform a bias correction to correct for any systematic offsets in $L_{\rm{C1}}$, $\beta$, $\gamma$, $L_{\rm{C2}}$ and $\delta$. \\
\\
The resulting median likelihood values (i.e. 50th percentile of the probability density function or PDF) are summarised in Table \ref{tab:bestfitvals}, together with uncertainties estimated from the 16th and 84th percentiles of the PDF. The SFR dependence that we observe ($\beta = 1.036^{+0.010}_{-0.040}$) is in agreement with what was observed in local galaxies ($d < 30$ Mpc; $\beta=1.10 \pm 0.06$), by \citet{2026Heesen}. However, the stellar mass dependence ($\propto M_*^\gamma$, with $\gamma=0.197^{+0.019}_{-0.016}$) is somewhat shallower than observed in previous works ($\gamma=0.43 \pm 0.01$ in \citetalias{Gurkan2018}, $\gamma=0.33\pm 0.04$ in \citetalias{S21}, $\gamma=0.330\pm 0.037$ in \citetalias{sdas2024}, and $\gamma=0.244 \pm 0.007$ in \citetalias{2026Shenoy}, to which our results are most similar), though the error bars on $\gamma$ are likely underestimated in previous works (see \citetalias{S21} and \citetalias{sdas2024} for details). The more sophisticated bias corrections applied in this work, particularly compared to \citetalias{S21} and \citetalias{sdas2024}, are also likely to contribute to the somewhat shallower dependence that we observe.\\
\\
To assess whether including an AGN component (equation \ref{agneq}) is preferred to the simpler SF-only model (equation~\ref{SFmodel}) when modelling our data, we use a Bayesian Information Criterion (BIC; \citealt{1978BIC}), where $\textrm{BIC} = k\ln(n) - 2\ln(\hat{\mathcal{L}})$ where $k$ is the number of free parameters, $n$ is the number of independent SFR and $M_*$ bins used in the fit and $\hat{\mathcal{L}}$ is the maximum likelihood. We find $\Delta$BIC$\,>\,$10 with a lower value of BIC for the model including an AGN component. Following the recommendation by \citet{BayesFactor}, this indicates that the model including an AGN component provides a significantly better fit than the model containing only a star-formation component.\\ 
\\
We plot the resulting model relations in Fig.~\ref{fig:mass_dep}. In the left panel, we use the median stellar mass in eight bins of SFR, calculate the radio luminosity using the bias-corrected median likelihood estimates of the parameters listed in Table~\ref{tab:bestfitvals}, and use a linear fit to show the resulting relation (in pink). In the central panel, we also show an indicative star-forming component and the AGN component separately: the SF component, shown by the black dashed line, is calculated using equation~\ref{SFmodel} at a median SFR value of 1\,$M_{\odot}\,\rm{yr}^{-1}$ or $\log_{10} (\psi/M_{\odot}~\rm{yr}^{-1}) = 0$. The AGN component is shown by the black dotted line and the grey shaded region is calculated using uncertainties on the value of $\delta$ and $L_{C2}$ as given in Table~\ref{tab:bestfitvals}. The pink line denotes the sum of the indicative SF component evaluated at $\log_{10}\psi = 0$ and the AGN mass dependence (equation \ref{agneq}). The low sSFR region is indicated by the yellow shaded region.\\
\\
To test for the possible influence of selection effects on our results, we repeated our analysis using a mass-complete sample, following the method of \cite{2010Pozzetti}. We re-analysed the data considering only the subset of galaxies with stellar mass greater than the 95 per cent stellar-mass completeness limit as a function of redshift (86,117 galaxies); our results remain consistent within the uncertainties (unsurprisingly, since the deviations from the canonical form occur at the high-mass end, largely unaffected by stellar-mass completeness limits). Similarly, we repeated the analysis within narrow fixed redshift bins (e.g. $\Delta z \sim 0.05$), finding that the trends remain (but with larger uncertainties). Lastly, we also repeated the experiment using radio flux densities measured at 113,228 random sky positions, finding that our results cannot be explained by some unknown aspect of the 150\,MHz data reduction. The main result is not driven by the 3.6~$\mu$m selection, since repeating the analysis on a stellar-mass-complete sample yields consistent results.
\renewcommand{\arraystretch}{1.3}
\setlength{\tabcolsep}{4pt}
\begin{table}
    \centering
    \caption{Bias-corrected median-likelihood estimates of the parameters of the two-component $L_{150\rm{MHz}}$--SFR model.}
    \label{tab:bestfitvals}
    \begin{tabular}{c l c l c}
        \hline
        \multicolumn{1}{c}{} & \multicolumn{2}{c}{\textbf{Star-formation component}} & \multicolumn{2}{c}{\textbf{AGN component}} \\
        \hline
        Normalisation &$L_{\rm C1}$ & $21.923^{+0.033}_{-0.024}$  &
        $L_{\rm C2}$ & $21.263^{+0.097}_{-0.178}$ \\[2pt]

        Mass dependence &$\gamma$ & $0.197^{+0.019}_{-0.016}$ &
        $\delta$ & $2.735^{+0.353}_{-0.273}$ \\[2pt]

        SFR dependence &$\beta$ & $1.036^{+0.010}_{-0.040}$ &
        --- & --- \\
        \hline
    \end{tabular}
\vspace{-0.2cm}
\end{table}

\section{Discussion}\label{discussion}
The observation that the minimum 150~MHz luminosity is independent of SFR for high stellar mass, low specific star formation rate galaxies is perhaps unsurprising, given that radio AGN are known to become increasingly common at higher stellar masses and typically associated with quiescent hosts \cite[e.g.][]{2019Sabater, 2023Best, drake2024, sdas2024, 2025Kondapally, 2025Hardcastle, 2026Arnaudova}. Indeed, it is also the case that the host's SFR makes an insignificant contribution to the radio luminosity of the most radio-loud AGN; however, we identify this behaviour for the first time in sources several orders of magnitude fainter ($L_\mathrm{150\,MHz} < 10^{22}$\,W\,Hz$^{-1}$), and after removing any detectable AGN activity from our sample.\\ 
\\
The steep dependence of the median 150\,MHz luminosity on stellar mass  ($\delta \sim 2.74$) suggests that -- if this is indeed due to widespread low-luminosity radio AGN activity -- the AGN radio luminosity increases more rapidly with host galaxy mass than expected from simple black-hole mass scaling. This may reflect a mass-dependent Eddington ratio distribution or AGN duty cycle (see e.g. \citealt{2018Bernhard}), such that more massive galaxies are more likely to host actively accreting black holes. This idea is consistent with the observed mass dependence of radio-loud AGN fractions,  $f_{\rm AGN} \propto M_*^{2.5}$, reported for low excitation radio galaxies  \citep{ 2012Janssen, 2022kondapally}, $f_{\rm RL} \propto M_*^{2.7}$ in the local galaxy population \citep{2015Williams} and $f_{\rm RL} \propto M_*^{2.1}$ reported for radio-loud AGN \citep{2008Shabala}, supporting the interpretation that the AGN contribution dominates the radio emission in high-mass galaxies. \\ 
\\
To investigate further, we estimate the Eddington-scaled accretion rate of black holes required to account for the observed radio luminosity in this high-mass regime. To do this, we calculate the Eddington ratio, $\lambda_{\rm Edd}$, following \cite{best2012}, and assume that the total energetic output of the black hole is equal to the jet mechanical power, $L_{\rm mech}$ (i.e. the bolometric radiative luminosity, $L_{\rm bol}$, is negligible). Under this assumption, $\lambda_{\rm Edd} \approx L_{\rm mech}/{L_{\rm Edd}}$. We estimate the mechanical jet power using the scaling relation of \citet{2010Cavagnolo} in the form used by \cite{best2012}:
\begin{equation}
\left( \frac{L_{\rm mech}}{\rm W} \right) = 7.3 \times 10^{36} 
\left( \frac{L_{1.4\,\rm GHz}}{10^{24}\,\rm W~Hz^{-1}} \right)^{0.7}
\end{equation}
where $L_{1.4~\rm GHz}$ is obtained from $L_{150~\rm MHz}$ assuming a spectral index $\alpha=0.7$. This should be regarded as a simplified approximation, since radio luminosity is a function of both jet power and time, evolving as the jet propagates and interacts with its environment (see e.g. \citealt{2018Hardcastle}). The Eddington luminosity is given by
\begin{equation}
\left( \frac{L_{\rm Edd}}{\rm W} \right)  = 1.3 \times 10^{31} 
\left( \frac{M_{\rm BH}}{M_\odot} \right) \rm,
\end{equation} 
where $M_{\rm BH}$ is estimated from the black hole mass--bulge mass relation of \cite{2004HaringRix}. In the central panel of Fig.~\ref{fig:mass_dep}, we show the relations corresponding to $\lambda_{\rm Edd} = 10^{-4}$ and $10^{-5}$ as blue dotted and dash-dotted lines, respectively. Taken at face value, these results suggest that the mass-dependent minimum radio luminosity is consistent with SMBHs accreting at effective Eddington-scaled accretion rates in the range $\lambda_{\rm Edd} \sim 10^{-4} - 10^{-5}$. However, since we have neglected any bolometric radiative output, the actual accretion rates may be higher (though, of course, we have removed any sources with identified radiative AGN activity from our sample). While these accretion rates are indicative only, given the large intrinsic scatter in each of the scaling relations employed, if we take them at face value, they are consistent with values obtained by recent studies of low-excitation radio galaxies (e.g. \citealt{2026Arnaudova}; Das et al. \textit{in prep}.), but again at orders of magnitude lower 150\,MHz luminosities. In the future the proposed ubiquity of low-accretion rate radio AGN in otherwise normal galaxies will be testable using brightness-temperature techniques on forthcoming very wide area sub-arcsecond LOFAR images from the iLOTSS survey (Shimwell et al. \textit{in prep}), building on the recent deep field works (e.g. \citealt{2022morabitoILT, 2024JurjendeJongILTElaisN1, 2024Ye}).

\subsection{Alternative explanations for the mass-dependent minimum radio luminosity}

It is possible that the low-level AGN activity that we observe could remain from previous higher Eddington ratio accretion episodes. However, remnants are found to comprise only a small fraction of the radio galaxy population, $\lesssim 10$ per cent \citep[e.g.][]{2018Mahatma, 2021Jurlin}, and modelling by \citet{2020Shabala} indicates that remnant lobes fade rapidly below the detection limit, with a dominant population of short-lived jets required to explain any remnant fraction above this level. We therefore consider it unlikely that remnant emission dominates the effect reported here.\footnote{We note, however, that the fraction of restarted sources is comparable to or higher than that of remnants \citep{2020Jurlin}, suggesting a rapid duty cycle in which jet activity resumes before the previous episode has faded. In this regime our inferred $\lambda_{\rm Edd}$ values are best interpreted as effective, time-averaged accretion rates rather than instantaneous ones.}\\
\\
It is also important to consider whether processes unrelated to black hole accretion could reproduce the observed trends. One possibility is enhanced cosmic-ray confinement in massive galaxies, where relativistic electrons may experience longer confinement times, thereby increasing synchrotron emission efficiency (e.g. calorimetric models such as those discussed by \citealt{2010LackiThompson}). This effect could, in principle, introduce a secondary dependence of radio luminosity on stellar mass - and we do see this for the SF component $(\gamma \sim 0.2)$. However, such models generally predict a smooth shift in the slope of the $L_{150\rm{MHz}}$--SFR relation, rather than the emergence of a distinct mass-dependent minimum radio luminosity at low sSFR. Moreover, the steep scaling we observe ($\delta \sim 2.74$) is significantly stronger than expected from simple calorimetric arguments \citep[e.g.][]{1989Voelk,1992Price, NickBeck1997}.

Star-forming galaxies are known to host low-mass X-ray binaries (LMXBs), associated with the old stellar populations, whose compact jets could in principle contribute synchrotron emission to the integrated radio continuum. However, the number and combined X-ray luminosity of LMXBs scale close to linearly with host stellar mass \citep{2004Gilfanov} with a modest secondary dependence on stellar age \citep{2012Zhang}. This linear dependence on stellar mass is far shallower than the $M_*^{2.74}$ dependence found here, making LMXBs an unlikely primary driver of the observed steepening and favouring an increasing AGN contribution instead. \\
\\
Environmental effects may also play a role: massive galaxies could undergo environment-quenching in dense environments (e.g. \citealt{2015Gabor, 2023Taylor}) such as groups and clusters, where interactions with the intracluster medium, including ram-pressure stripping (\citealt{2024Roberts}) or the presence of diffuse radio haloes, may enhance synchrotron emission \citep{2019VanWeeren}. While such effects could contribute to the observed scatter, they are unlikely to produce a coherent, monotonic scaling with stellar mass across the full sample. Nevertheless, an interesting avenue would be to systematically explore the role of environment using the wide-area imaging and spectroscopy provided by \textit{Euclid} \citep{2025Euclid} and spectroscopy from the WEAVE-LOFAR Survey \citep{2016Smithweave} in future work. \\
\\
Finally, systematic uncertainties in SFR estimates, particularly in quiescent galaxies, could mimic an apparent radio excess. In systems with low specific SFR, photometric SED fitting becomes increasingly sensitive to assumptions about star formation histories and dust heating from older stellar populations. This could result in underestimated SFRs and hence an artificial offset in the $L_{150\rm{MHz}}$--SFR relation. However, the persistence of the mass-dependent trend across different subsamples and its consistency with the known increase in AGN incidence with stellar mass (e.g. \citealt{2019Sabater, 2025Kondapally}) suggest that measurement uncertainties alone cannot account for the observed behaviour. We also compare the \textsc{Prospector} SFRs and stellar masses with \textsc{MAGPHYS} values (from \citetalias{S21}), finding offsets of $\Delta\log(\psi/M_\odot{\rm yr}^{-1})=0.14$ dex and $\Delta\log(M/M_\odot)=0.06$ dex, respectively. These differences are small compared to the intrinsic scatter ($\sigma_{\rm NMAD}= 0.17$ and 0.11 dex for SFR and stellar mass, respectively), and consistent with the typical differences between SED fitting codes, as discussed by e.g. \citep{2023pacifici}. Taken together, while these mechanisms may contribute at some level, none naturally explain both the amplitude and the steep stellar mass dependence of the observed radio excess. The most plausible interpretation is therefore that the mass-dependent floor arises from low-level AGN activity, which becomes increasingly prevalent in massive, quiescent galaxies and contributes significantly to the low-frequency radio emission in these very low luminosity sources.

\section{Conclusion}\label{conc}
In this work, we have used the deepest 150\,MHz observations in existence, covering the ELAIS-N1 deep field from the LOFAR Deep Fields DR2 \citep{2025Shimwell_LoTSSDR2Deep}, combined with multi-wavelength SED-derived galaxy properties from \citetalias{sdas2024}, to revisit the relationship between radio luminosity, star-formation rate, and stellar mass for a near-infrared selected sample of 113,228 galaxies. We recover the stellar-mass dependence of star-formation-driven radio emission as established in previous works (\citetalias{Gurkan2018, S21, sdas2024, 2026Shenoy}). However, the increased sensitivity of the DR2 radio mosaics enables us to probe the faint radio population in unprecedented detail. We find that at  $L_{150\rm MHz}\sim10^{20}$--$10^{22}$ W\,Hz$^{-1}$, where star-forming galaxies are expected to dominate, a minimum radio luminosity exists, which is independent of SFR and depends on stellar mass. This shows that the stellar mass dependence reported in previous works is not uniform across the sample but becomes increasingly important in galaxies with low sSFRs, where the observed radio luminosities exceed those expected from star formation alone. Assuming that low-level AGN accretion contributes to this radio emission, we find that the non-stellar radio emission is primarily driven by SMBHs accreting at varying effective Eddington-scaled accretion rates corresponding to $\lambda_{\rm{Edd}} \sim 10^{-5}-10^{-4}$, calculated using the relation from \cite{2010Cavagnolo}, assuming $L_{\rm{mech}} >> L_{\rm{bol}}$. We therefore propose a two-component model that takes into account not only star-formation-driven radio emission but also accretion-driven emission, such that:

\begin{equation}
\begin{aligned}
L_{150} ={}&
10^{21.923^{+0.033}_{-0.024}}
\left(\frac{\psi}{M_{\odot}\rm{yr}^{-1}}\right)^{1.036^{+0.010}_{-0.040}}
\left(\frac{M_*}{10^{10}M_\odot}\right)^{0.197^{+0.019}_{-0.016}}\\
& + 10^{21.263^{+0.097}_{-0.178}}
\left(\frac{M_*}{10^{11}M_\odot}\right)^{2.735^{+0.353}_{-0.273}}.
\end{aligned}
\end{equation}

\noindent Our results indicate that using a single-component $L_{150\rm MHz}$--SFR calibration can lead to systematic overestimates of SFR, especially in optically quiescent galaxies. These results have important implications for forthcoming deep radio surveys, including with LOFAR~2.0 and the Square Kilometre Array (SKA), which will probe larger numbers of more distant sources with progressively lower luminosities, where the low-level accretion-driven component identified here is most apparent.

\section*{Acknowledgements}
We sincerely thank the anonymous referee for their careful reading and insightful comments, which have significantly improved the quality of this manuscript. We thank Alyssa Drake for useful discussions. SS and DJBS acknowledge support from the UK Science and Technology Facilities Council (STFC) via the grant ST/X508408/1. DJBS and LRH acknowledge support from the UK STFC via grant ST/Y001028/1. DJBS and SD acknowledge support from the Leverhulme Trust via Research Project Grant RPG-2025-078. MIA acknowledges support from the UK STFC under grant ST/Y000951/1 and by NOIRLab, which is managed by the Association of Universities for Research in Astronomy (AURA) under a cooperative agreement with the U.S. National Science Foundation. KJD acknowledges support from the STFC through an Ernest Rutherford Fellowship (grant number ST/W003120/1). This research made use of the University of Hertfordshire high performance computing facility and the LOFAR-UK computing facility located at the University of Hertfordshire and supported by STFC (ST/V002414/1). LOFAR \citep{VanHaarlem13} is the Low Frequency Array designed and constructed by ASTRON. It has observing, data processing, and data storage facilities in several countries, which are owned by various parties (each with their own funding sources), and which are collectively operated by the ILT foundation under a joint scientific policy. The ILT resources have benefited from the following recent major funding sources: CNRS-INSU, Observatoire de Paris and Université d'Orléans, France; BMBF, MIWF-NRW, MPG, Germany; Science Foundation Ireland (SFI), Department of Business, Enterprise and Innovation (DBEI), Ireland; NWO, The Netherlands; The Science and Technology Facilities Council, UK; Ministry of Science and Higher Education, Poland; The Istituto Nazionale di Astrofisica (INAF), Italy. This work also made use of the following packages for Python: Astropy \citep{astropy:2013, astropy:2018, 2022astropy}, \texttt{Matplotlib} \citep{Hunter:2007}, \texttt{NumPy} \citep{harris2020numpy}, \texttt{SciPy} \citep{2020SciPy-NMeth}, \texttt{emcee} \citep{emcee}, and \texttt{pandas} \citep{reback2020pandas}. 

\section*{Data Availability}
The radio data used in this work are taken from the second data release of the LOFAR Deep Fields in the ELAIS-N1 field and can be found here: \url{https://lofar-surveys.org/deepfields_dr2.html}. SFRs and stellar mass estimates from \cite{sdas2024} used in this work are also publicly available at \url{https://lofar-surveys.org/deepfields_public_en1.html}.

\bibliographystyle{mnras}
\bibliography{refs}

\bsp	
\label{lastpage}
\end{document}